\documentstyle[aps,preprint,psfig]{revtex}

\begin{document}
\title{Interband Light Absorption at a  Rough Interface}
\author{Leonid S.~Braginsky\thanks{Electronic address: 
brag@isp.nsc.ru}}
\address{Institute of Semiconductor Physics, 630090, 
Novosibirsk, Russia}
\date{\today}
\maketitle
\maketitle
\begin{abstract}
Light absorption at the   boundary of 
indirect-band-gap and direct-forbidden gap semiconductors 
is analyzed.  It is found that the possibility of the electron 
momentum nonconservation at the interface  leads to essential 
enhancement of absorption in porous and microcrystalline 
semiconductors. The effect is more pronounced at a rough 
boundary due to enlargement of the share of the interface 
atoms. 
\end{abstract} 
\nopagebreak
\section{Introduction} 
\label{Introduction} 
It is well known 
that the direct interband electron transitions is the main 
mechanism of light absorption in pure semiconductors.  These 
transitions are direct owing to the momentum conservation low 
for the excited electron. The momentum, which this electron 
obtains from the light wave ($\sim \pi\hbar/\lambda_0$, where 
$\lambda_0$ is the wavelength of the light), is small in 
comparison with the electron momentum in the crystal ($\sim 
\pi\hbar/\lambda$, where $\lambda$ is the electron wavelength). 
It is clear, however, that the momentum is not conserved 
if the absorption takes place at the crystal boundary or at the 
interface between two crystals.  The possibility of indirect 
electron transitions at the interface  results in  
enhancement of the absorption.  This means that the 
considerable enhancement of light absorption has to be expected 
in porous and microcrystalline semiconductors where the share of 
the interface atoms is sufficiently large.

This interface mechanism of light absorption becomes most 
important if indirect electron transitions are more preferable 
than the direct ones. This happens, first, in indirect-band-gap 
semiconductors where the electron transition to the side valley 
in the bulk should be  accompanied by the electron-phonon  or  
electron-impurity  interaction.  Second, this happens  in 
direct-forbidden-band-gap semiconductors where the direct 
electron transitions between the top of the valence band and 
the bottom of the conduction band are prohibited.

Both possibilities are considered in this paper. We  investigate 
the frequency dependence of the absorption at the fundamental 
absorption edge and estimate the relative value of the interface 
absorption.
We consider also the interband light absorption at a rough 
interface. Such  interface is characteristic for  the
intercrystallite boundary in porous and microcrystalline 
semiconductors.

\section{The model}
We consider the light absorption at the  surface of the 
semiconductor quantum dot or the crystallite  embedded in 
an insulator media. Suppose that each size of the  
crystallite much exceeds the  
lattice constant.  On this assumption the envelope function 
approximation  is valid.  Moreover, this assumption allows us to 
consider the light absorption as  inelastic scattering of the 
electron at the interface if the electron wavelength is 
sufficiently small.  We introduce the Cartesian coordinate 
system, where the $z$ axis is normal to the interface and assume 
that semiconductor occupies the region $z<0$.  

The probability for the photon to be
absorbed in the crystallite is
\begin{equation}
\label{1}
\eta=\frac{(2\pi \hbar)^2e^2}{m_e^2c\omega n S}\sum_{\bf p,\ q}
\left|<f|\frac{\partial}{\partial z}|i>\right|^2
\delta(\varepsilon_c- \varepsilon_v-\hbar\omega),
\end{equation}
where $\hbar,\ e,\ m_e,$ and $c$  are the fundamental
constants, $\omega$ is the photon frequency,  $n$ is the
refraction index,  $S$ is the area
of the crystallite side $z=0$ where the absorption is considered, ${\bf
p}$ and ${\bf q}$ are  the electron momenta, $p$ and $q$ are their $z$
components, $\varepsilon_c({\bf q})$ and $\varepsilon_v({\bf p})$ are the
energies of the electron in the conduction and valence bands.
The electric field of the light is directed along the $z$ axis;
$i>$ and $f>$ are the wave functions of the electron before the
excitation (in the valence band) and after it (in the conduction
band) correspondingly. Wave functions $i>$ and $f>$  
are determined by the band structure of the crystallite  and 
the boundary conditions for the envelope wave functions at the 
interface between the crystallite and the insulator.

To
consider the Coulomb interaction, which occurs between the
exited electron and the hole in the valence band,  we have to 
input the weight factor $\Phi(\gamma)=\pi \gamma 
\exp\pi\gamma/(\sinh\pi\gamma)$ [where 
$\gamma=(qa_B/\hbar)^{-1}$, and $a_B$ is the effective Bohr 
radius] into the sum (\ref{1}) \cite{Elliot}.
The Coulomb interaction becomes important if  the electron and 
the hole are too close to the band extrema, so that their 
wavelengths are large.  In this case the  electron density 
at the place where the hole is situated is not small.  

It is possible to change summation in
Eq.~(\ref{1}) by integration over the electron energy  and
parallel-to-inter\-face components of the momentum
${\bf q_\parallel}$.  We obtain
\begin{eqnarray}
\label{2}
&&\eta=\frac{e^2m_cm_ha^2}{(2\pi)^2\hbar^4m_e^2\omega
cn}                                  \nonumber
\\   
&&\phantom{\eta=}\times
\int\frac{\Phi(\gamma)|{\cal{P}}_{vc}|^2\,d\varepsilon_c\,d^2
q_\parallel}
{\sqrt{2m_h(\hbar\omega-\varepsilon_c)-q_\parallel^2}\,
\sqrt{2m_c(\varepsilon_c-E_g)-q_\parallel^2}},\\
&&\mbox{where\ }\;\;\;
{\cal P}_{vc}=-i\frac{N}{N_s}<f|\frac{\partial}{\partial z}|i>,
\nonumber
\end{eqnarray}
$m_c$ and $m_h$   are the effective masses of the 
electron in the conduction and valence bands respectively, 
$N$ and $N_s$ are the numbers of atoms 
in the crystallite and at the interface, and $E_g$ is 
the gap.  The limits of integration are determined by the region 
where the expressions under the square roots in the integrand 
are positive.  

\section{Boundary conditions for the envelope wave function}
\subsection*{Boundary condition at the plane interface}
\subsubsection*{Interface of semiconductors with  nondegenerate 
band structure}

Let $z=0$ be the interface  between  semiconductor ($z<0$) 
and an insulator ($z>0$).  For a simple nondegenerate  electron 
spectrum the boundary conditions for the envelope wave function 
at the plane interface can be written in the form \cite{PR98}: 
\begin{eqnarray} \label{E1} 
&&\Psi_l(\tau_l^0)=b_{11}\Psi_r(\tau_{11}), \nonumber \\
&&b_{21}\Psi_l(\tau_{21})=\Psi_r(\tau_r^0).
\end{eqnarray}
Where $|b_{ik}|\sim 1$ and $|\tau_{ik}| \sim a$ are the 
parameters of the boundary conditions, which relate the 
envelopes at the {\it different} sites $\tau_{ik}$ at  
the interface. It is important that these sites 
are close to the interface.  The generally accepted form of the 
boundary conditions \cite{Ando} 
can be obtained from Eq.~(\ref{E1}) if we 
expand $\Psi(\tau)$ at the interface $z=0$ 
assuming $|\tau \Psi'| \ll 1\ ($where $\Psi'=\partial 
\Psi/\partial z)$.  Then we find
\[
\left(
\begin{array}{l}
\Psi_r \\ \Psi'_r
\end{array}
\right)=\hat{T}
\left(
\begin{array}{l}
\Psi_l \\ \Psi'_l
\end{array}
\right),
\mbox{ where }
\hat{T}=\|t_{ik}\|, 
\]
\begin{eqnarray}
\label{E2}
&&\begin{array}{ll}
t_{11}=\displaystyle\frac{b_{11}b_{21}\tau_{11}-\tau_r^0}
{b_{11}(\tau_{11}-\tau_r^0)}, &
t_{12}=\displaystyle 
\frac{b_{11}b_{21}\tau_{11}\tau_{21}-\tau_r^0\tau_l^0}
{b_{11}(\tau_{11}-\tau_r^0)}, \\   \ &\ \\
t_{21}=\displaystyle-\frac{b_{11}b_{21}-1}
{b_{11}(\tau_{11}-\tau_r^0)}, &
t_{22}=\displaystyle-\frac{b_{11}b_{21}\tau_{21}-\tau_l^0}
{b_{11}(\tau_{11}-\tau_r^0)}.
\end{array}
\end{eqnarray}                               
The values of $b_{ik}$, $\tau_{ik}$, and $t_{ik}$   are 
independent of the electron energy and can be used as the 
fenomenological parameters.  
The determinant of the $t_{ik}$ matrix should be equal to 
$m_r/m_l$ (where  $m_r$ and $m_l$ are effective masses of the 
electron on each side of the interface) in order to 
ensure the conservation of the electron probability flux at the 
interface.

It follows from Eq.~(\ref{E2}) that, in general, 
$t_{11}\sim 1$, $t_{12}\sim a$, $t_{21}\sim a^{-1}$, and 
$t_{22}\sim 1$. To understand the physical meaning of this 
estimation, let us consider the electron scattering at the 
interface. The envelope wave function of 
the electron is 
\[
\psi=\left\{
\begin{array}{ll}
[e^{ipz}+Re^{-ipz}]e^{i\bbox{p_\parallel \rho}}, & z<0,\\
Te^{-\gamma z+i\bbox{p_\parallel \rho}}, & z>0.
\end{array}
\right.
\]
Where ${\bf p}$ is the momentum of the electron, and $\gamma$ 
($p\ll\gamma\ll a^{-1}$) is the decay exponent of the electron 
wave function in the insulator.  We can use the boundary 
conditions (\ref{E2}) to obtain the reflection coefficient
\begin{equation}
\label{3} 
 1+R=-\frac{2ipt_{22}}{t_{21}+\gamma t_{11}-ipt_{22}}. 
\end{equation}
 The pole of this expression determines the 
energy $E$ of the electron level, which separates from the band 
at the interface. We can rewrite Eq.~(\ref{3}) as follows:
\begin{equation}
\label{4}
1+R=-\frac{2ip}{\kappa -ip},
\end{equation}
where $\kappa =\sqrt{2mE}=(t_{21}+\gamma t_{11})/t_{22}\simeq 
t_{21}/t_{22}$ is the decay exponent of 
the electron wave function of the interface state. Eq.~(\ref{4}) 
allows to relate the value of $t_{21}$ with the energy of the 
interface level. We see that the parameter $t_{21}$ indicates 
the value of  separation of the interface level apart off 
the band extremum. Indeed, for $t_{21}\sim a^{-1} \; E\sim 
\hbar^2/(2ma^2)$, i.e., the separation is of the order of the 
band width.  However, the interface level becomes close to the 
band extremum when $t_{21}\rightarrow 0$.

\subsubsection*{Interface of semiconductors with  degenerate 
band structure}

The boundary conditions for the envelope wave function becomes 
more complicated if the  intervalley or interband 
degeneracy occurs in the band structure of the semiconductor. 
For simplicity, let us consider the two-valley conduction band
presented on Fig.~1.  The boundary conditions for the 
electron in this band can be written as follows \cite{PR98}: 
\begin{eqnarray}
\label{31}
&&\Psi_1(\tau_1^0)=c_{11}\Psi_r(\tau_{11}), \nonumber \\
&&\Psi_2(\tau_2^0)=c_{22}\Psi_r(\tau_{22}), \\
&&c_{31}\Psi_1(\tau_{31})+c_{32}\Psi_2(\tau_{32})    
=\Psi_r(\tau_3^0).\nonumber 
\end{eqnarray}
Where parameters $c_{ik}\sim 1$ and $|\tau|\sim a$ are 
independent 
of the electron energy. The probability flux conservation holds  
at the interface  for arbitrary electron energies. To ensure 
this, the parameters of the boundary conditions (\ref{31}) must 
satisfy the following relations: 
\begin{eqnarray}
\label{32}
\frac{c_{11}(\tau_{11}-\tau_3^0)}{m_1}&=&
-\frac{c_{31}(\tau_{31}-\tau_1^0)}{m},\nonumber \\
\frac{c_{22}(\tau_{11}-\tau_3^0)}{m_2}&=&
-\frac{c_{32}(\tau_{31}-\tau_2^0)}{m}, \\
\tau_{11}&=&\tau_{22}.\nonumber
\end{eqnarray}
Where $m_1$ and $m_2$  are the effective masses of the 
electron in the central and side valleys of the conduction band 
of the semiconductor, and $m$ is the effective mass in the 
insulator.

An important simplification aries then the large band offset 
occurs at the contact of a two-valley semiconductor and   an
insulator.  If so, then $\Psi_r\propto\exp{(-\gamma_r z)}$, 
where the $\gamma_r$ value can be 
considered as independent of the electron energy. Eliminating 
$\Psi_r$ from Eqs.~(\ref{31}), we find 
\begin{eqnarray} 
\label{34} 
&&\Psi_1(\tilde{\tau}_{11})+\tilde{c}_{12}\Psi_2(\tilde{\tau}_{12})=0,\\
&&\tilde{c}_{21}\Psi_1(\tilde{\tau}_{21})+\Psi_2(\tilde{\tau}_{22})=0,
\nonumber 
\end{eqnarray}
where  $\tilde{c}_{ij}\sim c_{ij}$ and $\tilde{\tau}_{ij}\sim 
\tau_{ij}$ are known functions of $c_{ij}$, $\tau_{ij}$, and 
$\gamma_r$.  The boundary 
conditions (\ref{34}) looks quite alike Eq.~(\ref{E1}).  
Therefore, Eq.~(\ref{E2}) holds in this case too.
This allows us to use Eqs.~(\ref{E1}, 
\ref{E2}) to consider the light absorption also at the  
interface of indirect-band-gap semiconductor.

The more general boundary conditions that applicable at an 
interface with a large band offset has been proposed in 
Ref.~\cite{Volkov}. Ours, Eq.~(\ref{34}) holds in the 
effective mass approximation. This approximation has been used 
in \cite{PR98} to obtain Eqs.~(\ref{E1}, \ref{31}) and estimate 
the parameters $c_{ij}$ and $\tau_{ij}$.

\subsection*{Boundary conditions at a rough interface}

Nonlocal form of the boundary conditions (\ref{E1}) can be 
used to obtain the boundary conditions for the envelope wave 
function at a rough interface.  We  consider the special form 
of a rough interface that is presented on Fig.~2.  The interface 
looks like an array of the plane areas of the same 
crystallographic orientation. The random function 
$z=\xi(\bbox{r})$ of the  coordinates in $XY$ plane determines 
the positions of these areas relative to $z=0$.  

We  assume the average 
height of roughnesses $\sigma$ to be small in comparison with 
the electron wavelength. Then it is possible to describe the 
rough interface by means of the correlation function $W({\bf 
r'}, {\bf r''})=\overline{\xi({\bf r'})\xi({\bf r''})}$. 
For the homogeneous  rough interface $W({\bf 
r'}, {\bf r''})=W({\bf r'}-{\bf r''})$, i.e.,  the 
correlation function is the function of one variable:  
$\bbox{\rho}={\bf r'}-{\bf r''}$. There are two parameters that 
are most important when the  statistical properties of 
a rough interface is considered:  $\sigma^2=W(0)$ and the 
correlation length $l$ --- the mean attenuation length of the 
correlation function.  In our model the correlation length $l$ 
can be associated with the mean size of the plane area 
(Fig.~2). We shall analyze the relation between the light 
absorption and the parameters $\sigma$ and $l$.

The special form of the rough interface (Fig.~2) allows  us to 
apply the boundary conditions (\ref{E1}), which are applicable 
at a plane interface, at each plane $z=\xi$. Note that these 
boundary conditions are not applicable at the vicinity of the 
corner points (like point 1 on Fig.~2). We assume the mean size 
of the plane areas to be large in comparison with the lattice 
constant, so that the relative number of the corner points is 
small.

To obtain the boundary conditions at the rough 
interface, it is necessary to expand the envelopes in 
Eq.~(\ref{E1}) at  $z=\xi({\bf r})$, instead of $z=0$. 
This means that  $\bbox{\tau}$ values   in Eq.~(\ref{E2}) 
should be replaced by $\bbox{\tau}+\xi$, so that the matrix of 
the boundary conditions $\hat{T}$ becomes of the form 
\begin{equation}
\label{E3}
\hat{T}=
\left(
\begin{array}{ll}
t_{11}-t_{21}\xi\; &
t_{12}+(t_{11}-t_{22})\xi-t_{21}\xi^2 \\
t_{21} &
t_{22}+t_{21}\xi
\end{array}
\right),
\end{equation}
where $t_{ik}$ are the components of the boundary conditions  
matrix (\ref{E2}) at the plane interface.

It is important that now the boundary conditions matrix   
 depends on $\bbox{r}$. This results in the diffuse 
components of the wave functions. We write the envelopes as the 
sum of their average $\Phi_{l,r}$ and diffuse $\varphi_{l,r}$ 
components \cite{Bass}
\begin{equation}
\label{E4}
\psi_{l,r}=\Phi_{l,r}+\varphi_{l,r}, \mbox{ where \ \ } 
\overline{\psi_{l,r}}=\Phi_{l,r},\;\;\;
\overline{\varphi_{l,r}}=0.
\end{equation}

To obtain the boundary conditions for $\Phi_{l,r}$ and 
$\varphi_{l,r}$,  we substitute the envelopes  
(\ref{E4}) into Eq.~(\ref{E2}) using the boundary conditions 
matrix (\ref{E3}).  We also have  to average these equations and 
subtract the average equations from the initial ones. Then we 
obtain
\begin{eqnarray}
\label{E5}
&&\Phi_r=t_{11}\Phi_l+t_{12}\Phi'_l-t_{21}\overline{\xi^2}\Phi'_l
-t_{21}\overline{\xi\varphi_l}+(t_{11}-t_{22})\overline{\xi\varphi'_l},
\nonumber \\
&&\Phi'_r=t_{21}\Phi_l+t_{22}\Phi'_l+t_{21}\overline{\xi\varphi'_l},
 \\
&&\varphi_r=t_{11}\varphi_l+t_{12}\varphi'_l-t_{21}\xi\Phi_l
+(t_{11}-t_{22})\xi\Phi'_l-t_{21}\overline{\xi^2}\varphi'_l, \nonumber\\
&&\varphi'_r=t_{21}\varphi_l+t_{22}\varphi'_l+t_{21}\xi\Phi'_l.
\nonumber
\end{eqnarray}
The values $\xi^2-\overline{\xi^2}$, 
$\xi\varphi_l-\overline{\xi\varphi_l}$, and 
$\xi\varphi'_l-\overline{\xi\varphi'_l}$ are omitted; they are 
small if $\sigma/l\ll 1$.

We use Eqs.~(\ref{E5}) to consider the electron scattering 
at the rough interface. The envelope wave function can be 
written in the form:
\begin{eqnarray}
\label{E6}
&&\psi_l=e^{i({p^l_{\parallel}\rho}+p^l_{z}z)}
+Re^{i(p^l_{\parallel}\rho-p^l_{z}z)}, \nonumber \\
&&\psi_r=Te^{i(p^r_{\parallel}\rho+p^r_{z}z)},\nonumber \\
&&\varphi_l({\bf r})=\sum_{\bf k}\tilde{\varphi}_l({\bf k})
e^{i(k_{\parallel}\rho-k_{z}z)},  \\
&&\varphi_r({\bf r})=\sum_{\bf k}\tilde{\varphi}_r({\bf k})
e^{i(k_{\parallel}\rho+k_{z}z)}.  \nonumber\\
&&k_z=\sqrt{2m\varepsilon -{\bf k}^2}, \nonumber
\end{eqnarray}
where $\varepsilon$ is the electron energy.
It follows from Eq.~(\ref{E6}) that 
$\bbox{p^l_{\parallel}}=\bbox{p^r_{\parallel}}= 
\bbox{p_{\parallel}}$.  We use last two equations  
(\ref{E5}) to express the diffuse components $\varphi_l$ and 
$\varphi_r$ as functions of $\xi$, $R$, and $T$. The  
reflection ($R$) and transmission ($T$) coefficients  
can be obtained then from  first two equations (\ref{E5}). These 
equations accept the form of Eq.~(\ref{E2}) if we introduce the 
effective parameters $\tilde{t}_{ik}$ of the boundary conditions 
as follows: 
\begin{eqnarray} \label{E8} 
&&\tilde{t}_{11}=t_{11}+it_{21}\int\frac{k_z[t_{21}+ik_z(t_{11}-t_{22})]
}{t_{21}-ik_z(t_{11}+t_{22})-k_z^2t_{12}}
\tilde{W}({\bf k} -{\bf p}_\parallel)\,d^2{\bf k}, \nonumber \\
&&\tilde{t}_{12}=t_{12}-\sigma^2t_{21} 
+\int
\frac{[t_{21}^2+k_z^2(t_{11}-t_{22})^2]}
{t_{21}-ik_z(t_{11}+t_{22})-k_z^2t_{12}}
\tilde{W}({\bf k} -{\bf p}_\parallel)\,d^2{\bf k}, \nonumber \\
&&\tilde{t}_{21}=t_{21}-t_{21}^2\int\frac{k_z^2}
{t_{21}-ik_z(t_{11}+t_{22})-k_z^2t_{12}}
\tilde{W}({\bf k} -{\bf p}_\parallel)\,d^2{\bf k},  \\
&&\tilde{t}_{22}=t_{22}+it_{21}\int\frac{k_z
[t_{21}-ik_z(t_{11}-t_{22})]}
{t_{21}-ik_z(t_{11}+t_{22})-k_z^2t_{12}}
\tilde{W}({\bf k} -{\bf p}_\parallel)\,d^2{\bf k}. \nonumber 
\end{eqnarray}
Where $W(\bbox{\rho})=\overline{\xi(\bbox{r}+\bbox{\rho})
\xi(\bbox{r})}$ is the correlation function of the rough 
interface, and $\tilde{W}$ is its Fourier transform
\[
\tilde{W}({\bf k})=\int 
W(\bbox{\rho})e^{-ik\rho}\,d^2\bbox{\rho}.  
\]

For the diffuse component $\varphi_l({\bf r})$ we obtain
\begin{eqnarray}
\label{E10}
&&\varphi_l({\bf r})=\frac{2p^l_{z}}{(2\pi)^2}\int 
A(k_z)\xi({\bf k} - {\bf p}_\parallel)e^{i(k\rho-k_zz)}\,d^2{\bf 
k}, \mbox{ where }\\ 
&&A(k_z)= 
-\frac{k_zt_{21}(\tilde{t}_{22}-ip^r_{z}\tilde{t}_{12})
+(\tilde{t}_{21} -ip^r_{z}\tilde{t}_{11})
[k_z(t_{11}-t_{22})+it_{21}]}
{[t_{21}-ik_z(t_{11}+t_{22})-k_z^2t_{12}]
[\tilde{t}_{21}-ip^r_{z}\tilde{t}_{11}-ip^l_{z}\tilde{t}_{22}
-p^l_{z}p^r_{z}\tilde{t}_{12}]}.\nonumber
\end{eqnarray}

We should emphasize  two-fold influence of the interface 
roughness. First, the roughness of the interface results in 
the diffuse  component of the scattered wave. This 
component is always  small for the long wavelength 
electrons and vanish when $\lambda \rightarrow \infty$. 
Second, the effective parameters of the boundary conditions 
depend on the interface roughness. This effect depends on the 
relation between the electron wavelength and the correlate 
length of the rough interface. Indeed, we can assume 
$\tilde{W}({\bf k}-{\bf p}_\parallel) \sim\sigma^2\delta({\bf k}-{\bf 
k}_0)$ if $\lambda\ll l$. Then the corrections to the effective 
parameters of the boundary conditions  (\ref{E8}), coursed by 
the roughnesses, become proportional to $\sigma^2/(l\lambda)$ 
and vanish when $\lambda \rightarrow \infty$. 

Corrections to $t_{ik}$ become independent of $\lambda$ when 
$\lambda\gg l$. This is clear from Eq.~(\ref{E8}) because in 
this case  $|{\bf p}_\parallel|\ll|{\bf k}|$. It is easy 
to understand the reason. Formally, the roughnesses, each size 
of which is less then the electron wavelength, could be taken 
into account when the parameters $b_{ik}$ 
and $\tau_{ik}$  of the boundary conditions are obtained. In 
that consideration the roughnesses affect the value of these 
parameters but don't make them dependent on $\lambda$ 
\cite{PR98}. It follows from Eq.~(\ref{E8}) that $\lambda$ 
dependent corrections to $\tilde{t}_{ik}$ are of the order of 
$(l/\lambda)^2$; this is in agreement with estimations 
\cite{PR98}.

It should be noted that values of $\tilde{t}_{ik}$ are not real 
at a rough interface.  The reason has to do with the diffuse 
component of the scattered wave that spoils the time-inversion 
symmetry.

\newpage
\section{Interband light absorption at a plane interface}
\subsection*{Interface light absorption in 
direct-forbidden gap semiconductors}

Let us consider the light absorption at the interface between 
semiconductor ($z<0$) and an insulator ($z>0$). The band 
structure of the semiconductor is presented on Fig.~3.  We can 
write the electron wave functions as follows:

\begin{eqnarray}
\label{15}
&&i>=\frac{1}{\sqrt{N}}\left\{
\begin{array}{ll}
[v_p({\bf r})e^{ipz}+R_vv_p^*({\bf
r})e^{-ipz}]e^{i\bbox{\scriptstyle p_\parallel \rho}}, & z<0 \\
T_ve^{-\gamma_v z +i \bbox{\scriptstyle  p_\parallel \rho}}, & z>0,
\end{array}
\right.\nonumber \\
&& \ \\
&&f>=\frac{1}{\sqrt{N}}\left\{
\begin{array}{ll}
[u_q^*({\bf r})e^{-iqz}+R_cu_q({\bf
r})e^{iqz}]e^{i\bbox{\scriptstyle  q_\parallel \rho}}, & z<0 \\
T_ce^{-\gamma_c z +i \bbox{\scriptstyle  q_\parallel \rho}}, &
z>0.
\end{array} \right.\nonumber
\end{eqnarray}
Where    $u_q({\bf r})$ and $v_p({\bf r})$  are the Bloch
amplitudes in the conduction and valence band respectively;
$\gamma_{c.v}$ are decay exponents of the wave functions  
 apart off the interface; $R_v$, $R_c$, $T_v$, and 
$T_c$ are the reflection and transmission coefficients. It is 
possible to relate these values with positions of the interface 
levels (see Eq.~(\ref{4}))
\begin{equation}
\label{11.5}
1+R_v=-\frac{2ip}{\kappa_v -ip},\;\;\;\;
1+R_c=\frac{2iq}{\kappa_c +iq},
\end{equation}
where $\kappa_v= \sqrt{2m_hE_v}$, and $\kappa_c=
\sqrt{2m_c(E_g-E_c)}$.

In direct-forbidden gap semiconductors $u_0({\bf r})=0$.
We assume $u_q({\bf
r})=2i\sin(qa/2)\,w_q({\bf r})$, where $w_q({\bf r})$ is the
periodic function of ${\bf r}$ that is not vanished at $q=0$.
Then for the  interband matrix element ${\cal P}_{vc}$ 
we obtain
\begin{eqnarray}
\label{7b}
&&{\cal P}_{vc}=\frac{N_s}{N}(1+R_v)(1+R_c^*) U
\,\delta_{\bbox{\scriptstyle{p_\parallel}},
\bbox{\scriptstyle{q_\parallel}}}, 
\mbox{ where}\nonumber\\
&&U=\int_{\Omega_0}
w_0\frac{\partial v_0}{\partial z}\,d^3{\bf r},
\mbox{ so that}\\
&&|{\cal 
P}_{vc}|^2=\frac{4U^2N_s^2}{N^2}
\frac{(\varepsilon_c-E_g)(\omega-\varepsilon_c)}
{(\varepsilon_c-E_c)(\omega-\varepsilon_c+E_v)}
\,\delta_{\bbox{\scriptstyle{p_\parallel}},
\bbox{\scriptstyle{q_\parallel}}} \nonumber.
\end{eqnarray}
We see that energy dependence of $|{\cal P}_{vc}|^2$ is 
sensitive to the position of the interface levels $E_v$ and 
$E_c$, whether or not they are close to the corresponding band 
extrema.  If we assume the Coulomb exponent $\Phi(\gamma)$ as 
independent of q, then the result of integration (\ref{2}) can 
be written as follows:  
\begin{eqnarray}
\label{10b}
&&\eta\propto (\hbar\omega-E_g)^{\nu},
 \mbox{ where}  \\
&& \nu=\left\{
\begin{array}{ll}
3&\mbox{ if $\;\; \hbar\omega-E_g\ll \min(E_v,\ E_g-E_c)$},\\
2&\mbox{ if $\;\; \min(E_v,E_g-E_c)\ll \hbar\omega-E_g \ll
\max(E_v,E_c)$},\nonumber\\
1&\mbox{ if $\;\; \hbar\omega-E_g\gg \max(E_v,\ E_g-E_c)$}.\\
\end{array}
\right.
\end{eqnarray}

\subsection*{Interface light absorption in indirect-band-gap 
semiconductors}

The band structure of the semiconductor is presented on 
Fig.~1.  For simplicity, we assume the nondegenerate valence 
band. There are two valleys in the conduction band: the central 
valley with the minimum at the center of  Brillouin zone, and 
the side valley at the edge of it. It is 
important that the side valley of the conduction band is 
situated in the direction of the normal to the interface.

 We  write the electron wave functions as
follows:
\begin{eqnarray}
\label{10}
&&i>=\frac{1}{\sqrt{N}}
\left[ v({\bf r})e^{ipz}+R_vv^*({\bf 
r})e^{-ipz}\right]e^{i\bbox{p_\parallel \rho}},  
\\ 
&&f>=\frac{1}{\sqrt{N}} 
\left[ R_cu_c({\bf r})e^{\kappa 
z}+R_su_s^*({\bf r})e^{i\left(q-\frac{\pi}{a}\right)z}+ u_s({\bf 
r}) e^{-i\left(q-\frac{\pi}{a}\right)z}\right] 
e^{i\bbox{q_\parallel \rho}}. 
\nonumber
\end{eqnarray}
\noindent Where  $v$, 
$u_c$, and $u_s$ are the Bloch amplitudes in the
valence band, the central  and side valleys of the conduction 
band; $p$, $\kappa$, and $q$ are  $z$ components of the wave 
numbers  in these valleys.    The coefficients 
$R_v,\ R_c$, and $R_s$ are determined by the 
boundary conditions for the envelopes.

For the interband matrix element ${\cal P}_{vc}$ we 
find \cite{PR98}
\begin{eqnarray}
\label{12}
&&{\cal P}_{vc}=\frac{4pq}{(\kappa_c+iq)(\kappa_v-ip)}
\left[ \frac{CP_c}{a}\cdot\frac{\kappa+\kappa_v}{\kappa^2+p^2}
+\frac{1}{2}P_s \right],\\
&&P_{c}=-i\hbar\int_{\Omega_0} u_0^*\frac{\partial
v}{\partial z}\,d^3r,\ \ \
P_{s}=-i\hbar\int_{\Omega_0} u_q^*\frac{\partial
v}{\partial z}\,d^3r, \nonumber
\end{eqnarray}
 $\varepsilon_c$ and $\varepsilon_v$ are the electron 
energies in the conduction and valence bands respectively, $C$ 
is the coefficient depended on the interface structure, and 
$\Omega_0$ is the unit cell.

The energy dependence of  $|{\cal{P}}_{vc}|^2$ (for $p\ll 
\kappa$) is the same as Eq.~(\ref{7b}), therefore
\begin{equation} 
\label{20} 
\eta\propto 
(\hbar\omega-E_g)^{\nu} \left[|P_{s}|^2+\beta 
\frac{|P_{c}|^2}{(\kappa a)^2}\right], 
\end{equation}
where $\nu$ is determined by Eq.~(\ref{10b}), and $\beta$ is the
dimensionless parameter, which has been aroused from the first 
term of the expression (\ref{12}).  Two terms in the 
square brackets of Eq.~(\ref{20}) can be interpreted as follows: 
the first one corresponds to the immediate  transition of the 
electron to the side valley at the interface,  second term  
corresponds to the excitation of the electron to the virtual 
interface state of the central valley  subsequented by  
conversion to the side valley, it  prevails if the valleys 
minima are close ($\kappa a\ll 1$).

The Coulomb interaction between the electron in the conduction 
band and hole in the valence bands can affect the frequency 
dependence of the absorption.  Indeed, 
$\Phi(\gamma)$ is independent of $q$ only if $qa_B\gg \hbar$.  
Otherwise, if $qa_B\ll \hbar$, it is proportional to $q^{-1}$; 
in this case  the exponent $\nu$ changes, so that $\nu 
\rightarrow \nu-1/2$ when $\hbar\omega -E_g \ll \mu 
e^4/(2\epsilon^2\hbar^2)$ (where $\mu^{-1}=m_c^{-1}+m_h^{-1}$, 
$\epsilon=n^2$).

Thus, the frequency dependence of the absorption    at the 
fundamental absorption edge essentially depends on the conditions at 
the interface. Namely, whether or not the interface electron 
levels are close to the corresponding bands extrema.

\section{Light absorption at a rough interface}

According to Eq.~(\ref{E4}), we  write
\begin{equation}
\label{E13}
\psi_i=\Psi_i+\varphi_i\;\;\;\mbox{and}\;\;\;\psi_f=\Psi_f+\varphi_f,
\end{equation}
where $\Psi_{i,f}$ and $\varphi_{i,f}$ are the average and 
diffuse components of the corresponding wave functions:
\begin{eqnarray*}
&&\Psi_i=\frac{1}{\sqrt{N}}\left\{
\begin{array}{ll}
[v_p({\bf r})e^{ipz}+R_vv_p^*({\bf 
r})e^{-ipz}]e^{i\bbox{\scriptstyle p_\parallel \rho}}, & z<0 \\ 
T_ve^{-\gamma_v z +i \bbox{{\scriptstyle  p_\parallel \rho}}}, 
& z>0, 
\end{array} 
\right.  \\
&&\varphi_i({\bf r})=\frac{2p}{(2\pi)^2\sqrt{N}}\int 
v_k({\bf r})A_v(k_z)\xi({\bf k} - {\bf 
p}_\parallel)e^{i(k\rho-k_zz)}\,d^2{\bf k}
\end{eqnarray*}
\begin{equation}
\label{E14}
\end{equation}
\begin{eqnarray*}
&&\Psi_f=\frac{1}{\sqrt{N}}\left\{
\begin{array}{ll}
[u_q^*({\bf r})e^{-iqz}+R_cu_q({\bf 
r})e^{iqz}]e^{i\bbox{\scriptstyle  q_\parallel \rho}}, & z<0 \\ 
T_ce^{-\gamma_c z +i \bbox{\scriptstyle  q_\parallel \rho}}, & z>0, 
\end{array}
\right.     \\
&&\varphi_f({\bf r})=\frac{2q}{(2\pi)^2\sqrt{N}}\int 
u_k^*({\bf r})A_c(k_z)\xi({\bf k} - {\bf 
q}_\parallel)e^{i(k\rho+k_zz)}\,d^2{\bf k},
\end{eqnarray*}
where $A_v$ and $A_c$ are the coefficients $A(k_z)$ (\ref{E10}) 
for the valence and conduction bands respectively.
By substituting the wave functions (\ref{E14}) into 
Eq.~(\ref{15}),  we express the matrix element  ${\cal P}_{vc}$ 
as follows:
\begin{eqnarray}
\label{E15}
&& {\cal P}_{vc}={\cal P}^{(1)}_{vc}+{\cal P}^{(2)}_{vc}+{\cal 
P}^{(3)}_{vc},\;\;\;\;\mbox{where}
\nonumber \\
&&{\cal P}^{(1)}_{vc}=\int{\psi_f^*\hat H_{\rm int} \psi_i\,dV},
\\&&
{\cal P}^{(2)}_{vc}=\int[\psi_f^*\hat H_{\rm int} \varphi_i+
\varphi_f^*\hat H_{\rm int} \psi_i]\,dV, \nonumber \\
&&\mbox{and}\;\;\;
{\cal P}^{(3)}_{vc}=\int{\varphi_f^*\hat H_{\rm int} 
\varphi_i\,dV}. \nonumber
\end{eqnarray}
We have to obtain the square module of Eq.~(\ref{E15}) and then 
to average it over the realizations of the random function 
$\xi(\bbox{\rho})$. For $\overline{|{\cal P}_{vc}|^2}$ we have
\[
\overline{|{\cal P}_{vc}|^2}=
|{\cal P}^{(1)}_{vc}|^2+\overline{|{\cal P}^{(2)}_{vc}|^2}+
{\cal P}^{(1)}_{vc}\overline{{\cal P}^{(3)^*}_{vc}}+
{\cal P}^{(1)^*}_{vc}\overline{{\cal P}^{(3)}_{vc}},
\]
  From (\ref{E14})  and (\ref{E15}) we 
obtain
\begin{mathletters}
\label{21}
\begin{equation}
\label{E16}
{\cal P}^{(1)}_{vc}=\frac{N_s}{N}(1+R_v)(1+R_c^*) U
\,\delta_{\bbox{\scriptstyle{p_\parallel}}, 
\bbox{\scriptstyle{q_\parallel}}},
\end{equation}
\begin{eqnarray}
\label{E17}
&&\overline{|{\cal 
P}^{(2)}_{vc}|^2}=\frac{4N^2_s|U|^2}{N^2S}W({\bf 
p}_{\parallel}-{\bf q}_{\parallel})\\
&&\phantom{\overline{|{\cal 
P}^{(2)}_{vc}|^2}}\times
\left|p(1+R^*_c)A_v(k_z)-q(1+R_v)A_c^*(k_z)\right|^2,
\nonumber
\end{eqnarray}
and
\begin{equation}
\label{E18}
\overline{{\cal 
P}^{(3)}_{vc}}=4ipqU\frac{N_s}{N}
\delta_{\bbox{\scriptstyle{p_\parallel}}, 
\bbox{\scriptstyle{q_\parallel}}}
\int A_v(k_z)A_c^*(k_z)
W({\bf k}_{\parallel}-{\bf q}_{\parallel})\,d^2{\bf k}.
\end{equation}
\end{mathletters}

The expressions (\ref{21})  allow to estimate 
influence of the roughness on the value and the frequency 
dependence of the absorption. Note that 
the ${\cal P}^{(1)}_{vc}$ value is of the same form as that for 
the plane interface. The main difference concerns parameters 
of the  boundary conditions  Eq.~(\ref{E8}), wherein an 
influence of the interface roughness has been taken into 
account. This results in the shift of the interface levels that 
manifests itself in the frequency dependence of the absorption.  
This effect is most important if the interface electron levels 
are close to both, valence and conduction, bands or if the 
interface is smooth; the values of $1+R_v$ and $1+R_c$ are not 
small in these cases. 

Indeed, the expressions (\ref{E17}) and (\ref{E18}) 
determine an influence of the diffuse components of the 
scattered waves on the absorption; the values of these 
components are as small as $\sigma/\lambda$. On the contrary, 
this small parameter is absent in Eq.~(\ref{E16}) that 
determines an influence of the average component (i.e., position 
of the interface electron level) on the absorption. However, 
Eq.~(\ref{E16})  is proportional to $(1+R_v)(1+R^*_c)$. This 
value is small if the interface level is not close to any band. 
Then the values of (\ref{E17}) and (\ref{E18}) may be of the 
same order or even exceed (\ref{E16}).

Comparing (\ref{E17}) and (\ref{E18}) with (\ref{E16}), we find 
that the roughness influence on the absorption is determined by 
the  value
\begin{eqnarray}
\label{chi}
&&\chi =
\frac{pq\sigma^2}{|1+R_c||1+R_v|}\sim 
\frac{\sigma^2}{4}\sqrt{(\kappa_v^2+p^2)(\kappa_c^2+q^2)}
\nonumber \\
&&\phantom{\chi}\sim
\frac{\sigma^2}{2}\sqrt{m_cm_h(\hbar\omega+E_v-E_g)
(\hbar\omega-E_c)}.
\end{eqnarray}

The correlation length $l$ of the rough interface is also 
important for the light absorption. We can assume 
$\tilde{W}({\bf k})=\sigma^2\delta({\bf k})$  if $l\gg \lambda$. 
It can be shown that $\overline{|{\cal P}^{(2)}_{vc}|^2}+ {\cal 
P}^{(1)}_{vc}\overline{{\cal P}^{(3)^*}_{vc}}+ {\cal 
P}^{(1)^*}_{vc}\overline{{\cal P}^{(3)}_{vc}}=0$ in this case. 
It is easy to understand the reason.  Roughnesses,  the mean 
length of which (Fig.~2) essentially exceeds the electron 
wavelength, couldn't affect the electron properties of the 
interface.

In the opposite limiting case $l\ll \lambda$ the roughness 
influence is determined by the term ${\cal P}^{(3)}$. 
Enhancement of the absorption in this case arises due to 
an increase of the number of the interface atoms $N_s\rightarrow 
N_s\sigma/a$, in vicinity of which the interband absorption with 
the momentum nonconservation occurs.

The interesting situation arises when $l\sim \lambda$. In this 
case the diffuse scattering that is described by  the term 
$\overline{|{\cal P}_{vc}^{(2)}|^2}\sim \kappa_c^2\sigma^2 
l^2[2m_{c\parallel}(\hbar\omega-E_g)]|{\cal P}^{(1)}_{vc}|^2$
 leads to the change in the frequency dependence 
of the absorption. 
This means the rapid increase of absorption at  the 
high frequencies  when $2m_{c\parallel}(\hbar\omega-E_g)l^2\sim 
1$, so that the exponent  $\nu$ changes its value from 
$\nu\rightarrow \nu+1$.

We considered the interband light absorption at the interface 
and found that possibility of the electron momentum 
nonconservation leads to enhancement of the absorption in the 
small microcrystallite.  To estimate the value of the interface 
absorption, we have to compare the absorption of the 
microcrystalline solid composed from the crystallites under 
consideration, $\alpha=\eta/L\propto w/L$ , with the light absorption in the 
bulk semiconductor. Here  $L$ is  
length of the crystallite under consideration, $w$ is the 
main size of the region at the interface where the interband 
electron transitions with the momentum nonconservation are 
possible. To estimate this value, we can use Eqs.~(\ref{11.5}, 
\ref{7b}): $w\sim \lambda$ if the interface 
electron levels are close to both (conduction and valence) bands 
extrema, $w\sim \sqrt{a\lambda}$ if the interface level is close 
to any (conduction or valence) band extremum, and $w\sim a$ if 
both interface levels separate too far off the band extrema. 
The small value $w/L$ characterizes the ratio of the number of 
atoms at this interface region to the number of atoms in the 
whole of solid. 

 In order to the interface absorption becomes important in Ge and
the indirect-band-gap semiconductors of the A$_{\rm III}$B$_{\rm 
V}$ group this small value should exceed the small parameter of 
electron-phonon interaction $g$ (this value is about 
10$^{-3}$--10$^{-2}$). In these semiconductors the interface 
absorption   essentially increases due to the intervalley 
conversion, which is determined by the second term in 
Eq.~(\ref{20}). That is possible if $w/L(\kappa a)^{-2}\geq g$.

The main mechanism of light absorption in bulk Si  is 
the impurity absorption. In order to the interface absorption 
becomes significant in this material, the number of the 
interface atoms should exceed the number of impurities. 
It seems possible that considered here indirect interband 
electron transitions at the interface are responsible for an 
increase of luminescence of the porous Si (see \cite{Lockwood} 
and references therein).

Enhancement of the light absorption at the fundamental 
absorption edge has to be expected in direct-forbidden gap 
semiconductors, wherein the direct electron transitions between 
the band extrema are prohibited. The essential difference in the 
effective masses of the valence and conduction bands is 
favourable for the effect. 
Such situation is characteristic for transition metal 
oxides semiconductors \cite{WO3}. The conduction band in these 
materials composed mainly of $d$ orbitals of the metal.  These 
orbitals are strongly localized.  For this reason the conduction 
band  is narrow and become even flat in certain 
direction ($\Gamma$-$M$ for TiO$_2$). It is important 
that direct interband electron transition is dipole-forbidden in 
this material.  There is two-fold advantage from the indirect 
electron transitions.  First, they are allowed, i.e., the dipole 
matrix element of the indirect transitions is not small. Second 
advantage arises due to the large density of states for the 
electron in the conduction band.

Let us compare the
interband transitions in the bulk of semiconductor $1\rightarrow 
2$ and at the interface $3\rightarrow 4$ (Fig.~3). The 
absorption is proportional to the density of electron states in 
each band.  The  density of states in the conduction band of 
TiO$_2$ essentially exceeds that  in the valence band.  
Nevertheless, it is the density of electron states in the 
valence band which determines the absorption in the bulk of 
crystal.  This happens because of the momentum 
conservation law, which makes the electron states with the large 
momentums inaccessible for the exited electrons. On the 
contrary, indirect electron transitions make such states 
accessible.

We can use  Eq.~(\ref{10b}) to estimate  the 
value of absorption coefficient  of the microcrystalline 
direct-forbidden gap semiconductor:
\[
\frac{\alpha}{\alpha_0}\sim
\frac{w}{L}\,\left(\frac{m_c}{m_h}\right)^2
\sim \frac{w}{L}\,\left(\frac{W_v}{W_c}\right)^2,
\]
where $\alpha_0$ is the absorption coefficient of the bulk
monocrystal,  $W_v$ is width of the 
valence band 
($W_c\sim 10\,$meV and $W_v\simeq 1\,$eV  for TiO$_2$ 
\cite{WO3}).  Thus, the interface mechanism of the absorption 
becomes comparable with the bulk one for the microcrystalline 
TiO$_2$ the mean size of the crystallite in which is $L\leq 
w(W_v/W_c)^2\sim 100\,$nm.  Significant increase of optical 
absorption has been observed in the microcrystalline BaTiO$_3$ 
\cite{Yamada}.

We found that the absorption value and its frequency dependence 
at the fundamental absorption edge are sensitive to the 
structure of the interface.
Thus, essential increase of the absorption has to be expected at 
a smooth interface. For such interface the main size of the 
interface region (i.e., the region at the interface where one 
crystal structure changes into another)  much exceeds the lattice 
constant. It is found that $t_{21}=0$ at a smooth 
interface \cite{Ivchenko}, so that $\kappa_c=\kappa_v=0$, and 
the interband matrix element ${\cal P}_{vc}$ increases, because 
it no longer proportional to the small ratio $(a/\lambda)^2$ 
[see Eqs.~(\ref{11.5}, \ref{7b})].  Indeed, the significant (in 
100 times) enhancement of the light absorption has been observed 
at the interface a-Si/mc-Si~\cite{Beck}.

The possible existence 
of the interface electron levels is essential for the optical 
properties of a sharp interface.  Energy position of these 
levels depends not only on the bordering materials, but also on 
the interface itself.  Structure of the interface as well as 
impurities and defects on it affect positions of these levels.  
Their positions can be measured in optical experiments as the 
singularities in the absorption spectrum at the energies below 
the gap value.

It seems  that the electron interface level should be
close to the valence band at least  in  wide-gap semiconductors.
The interface level becomes empty then it is shifted too far
off the top of the valence band.  This results in a large
surface charge and a strong band bending that is not favorable
from the energetical point of view.  Nevertheless, the interface
level can be shifted as the result of structure reconstruction
of the interface. This reconstruction does not essentially
affect the interatomic spaces or angles, but it makes the
interface level to be closer to the top of the valence band.

The roughness of the interface is one of the possible ways of
such  reconstruction. It follows from our
consideration Eq.~(\ref{E8}) that significant shift of
the interface level occurs if the correlation length of
rough interface is small ($\kappa_v l \ll \hbar$). That
causes the interface level to be closer to the band
extremum.  This is the particular case when the structure
reconstruction of the interface decreases the interface energy.
If so, then the rough interface becomes more favorable than the
plane one.

It seems strange that the interface roughnesses result in 
only the shift of the interface levels, but they can't be the 
origin of such levels. In other words, $\tilde{t}_{21}$ in 
Eq.~(\ref{E8})  vanishes for an appropriate correlation 
function $\tilde{W}(k)$, but it always equal to zero if 
$t_{21}=0$.  This is the result of our assumption that the 
interface is smooth ($l\gg a$).  Perhaps, the interface level 
can be separated from the band extremum due to the interface 
roughnesses if $l\sim a$. In that case the interface level 
arises as the result of the rearrangement of the chemical bonds 
at the rough interface.

The value of $\chi$ Eq.~(\ref{chi}) determines the roughnesses 
influence on the absorption. This value is small, $\chi\sim 
\sigma^2/\lambda^2$, if the interface electron levels are close 
to both bands extrema. In this case the main size $w$ of the 
region at the interface where the momentum nonconservation is 
possible is of the order of the electron wave length $\lambda$, 
so that the small roughnesses can't affect the absorption. This 
is not the case if both interface levels shift far apart of the 
bands extrema, then $w\sim a$ and $\chi$ value is of about or 
even exceeds unity.

The roughnesses influence on the absorption is
significant at the low frequencies when $\hbar\omega\leq E_g +
\hbar^2(2m_cl^2)^{-1}$; enhancement of absorption in this case
is due to an effective increase of the number of the interface
atoms ($N_s\rightarrow N_s\sigma/a$) in the vicinity of which the
interface light absorption occurs (Fig.~4).

In conclusion, we show that the possibility of the momentum
nonconservation at the interface  leads to enhancement of 
the interband light absorption in small  crystallites, the size 
of which is about a few $10\,$nm.  The interband absorption is 
sensitive to the interface electron levels; namely, whether or 
not they are close to the bands extrema. The influence of the 
interface roughnesses is essential if these levels are not close 
to any or both conduction and valence bands.  The effect of the 
roughnesses is two-fold. First, they result in the diffuse  
electron scattering at the interface. This leads to enhancement 
of the absorption due to the effective increase of the share of 
the interface atoms. Second, roughnesses result in effective 
shift of the electron interface levels.


\subsection*{Acknowledgments}
The author  wishes to thank Prof.\ E.\ Ivchenko, who address him 
to influence of the interface roughnesses on the absorption, 
Prof.\ V.\ Volkov and Prof.\ M.\ Entin for the helpful 
discussions.  This work was supported by the  Russian Foundation 
for the Basic Research, Grant No.~99-02-17019.



\begin{figure}
\centerline{\psfig{figure=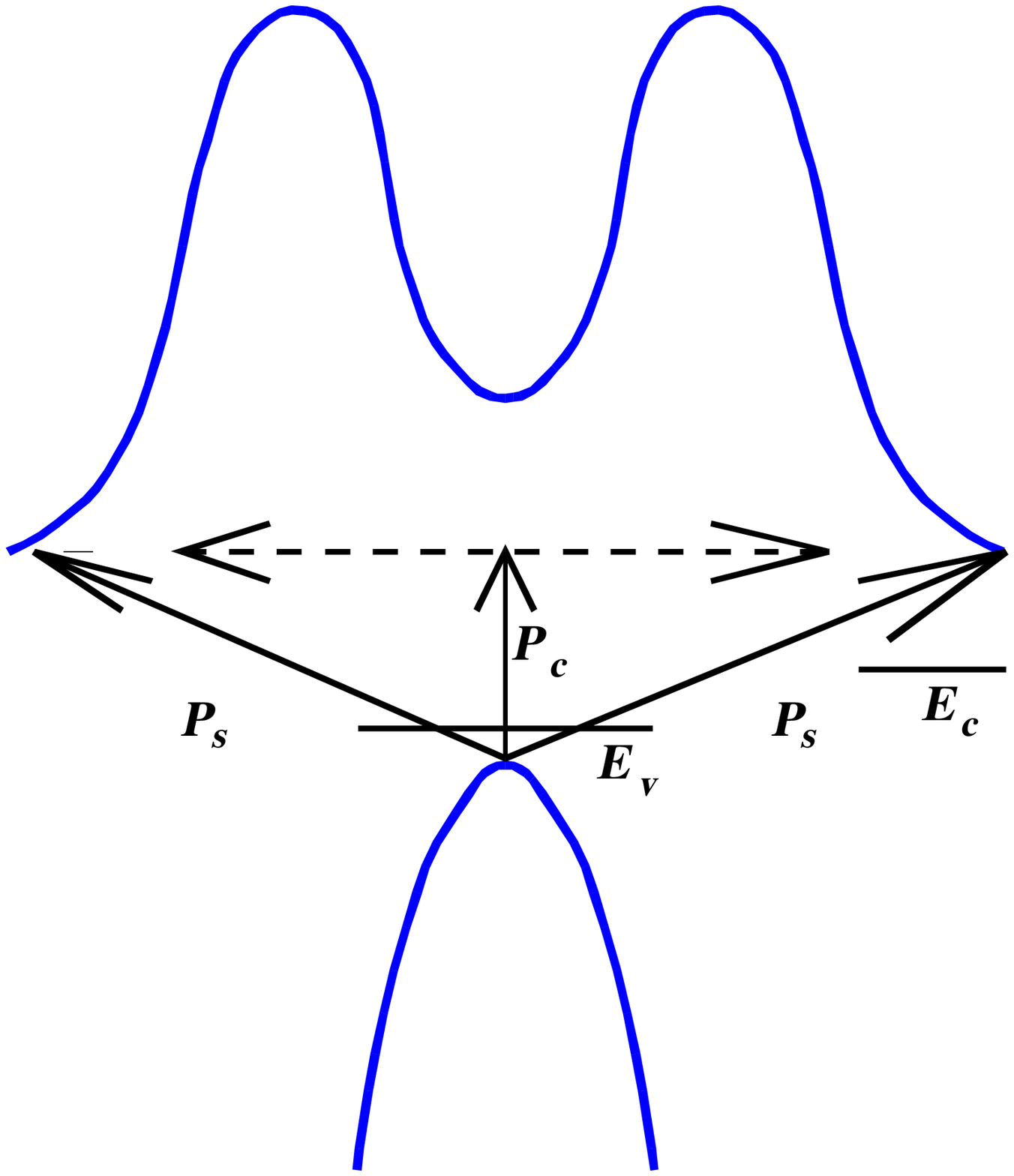,height=6.0in}}
\caption{
The two ways of the light absorption at the 
indirect-band-gap semiconductor interface: the immediate electron 
transition to the side valley $P_s$, and  the  vertical 
transition $P_c$ followed by the conversion to the side valley 
(dotted arrow).   $E_v$ and 
$E_c$ are the interface levels. 
}
\end{figure}

\begin{figure}
\centerline{\psfig{figure=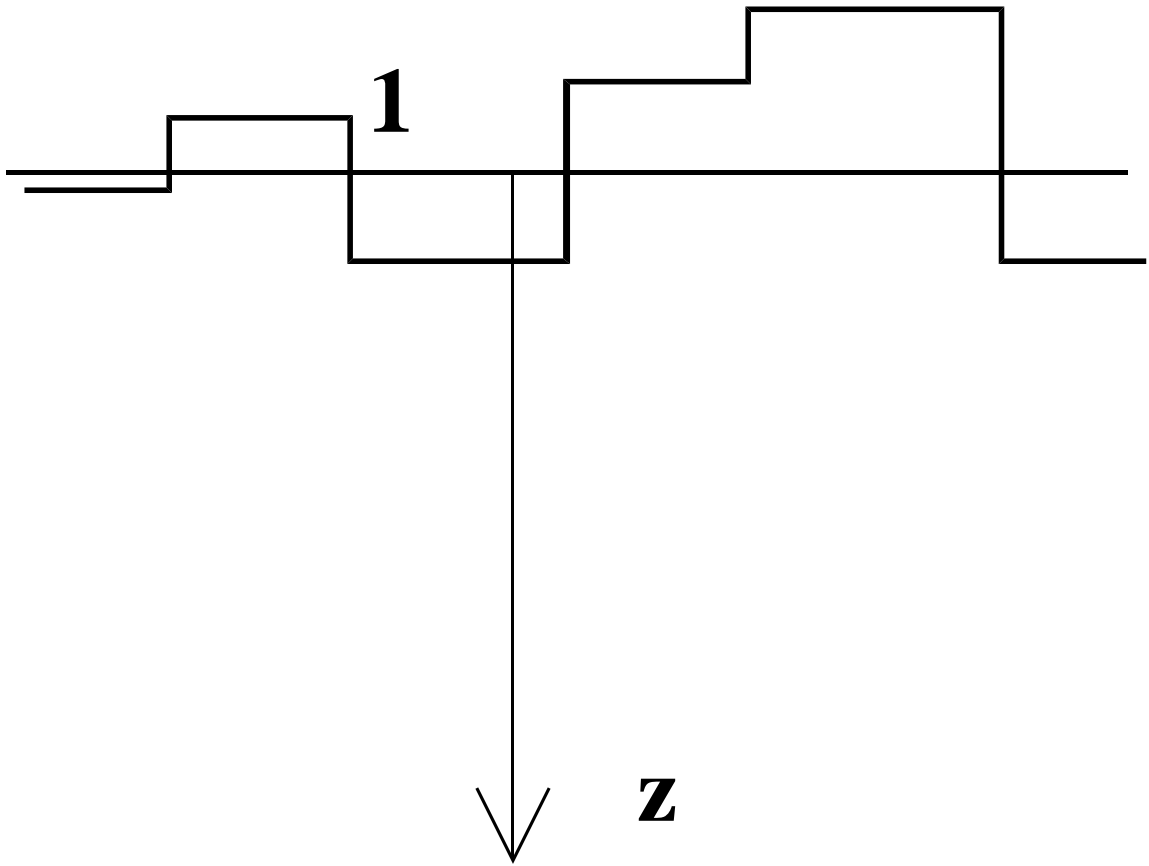,height=6.0in}}
\caption{
The model of the rough interface. Side view.
}
\end{figure}

\begin{figure}
\centerline{\psfig{figure=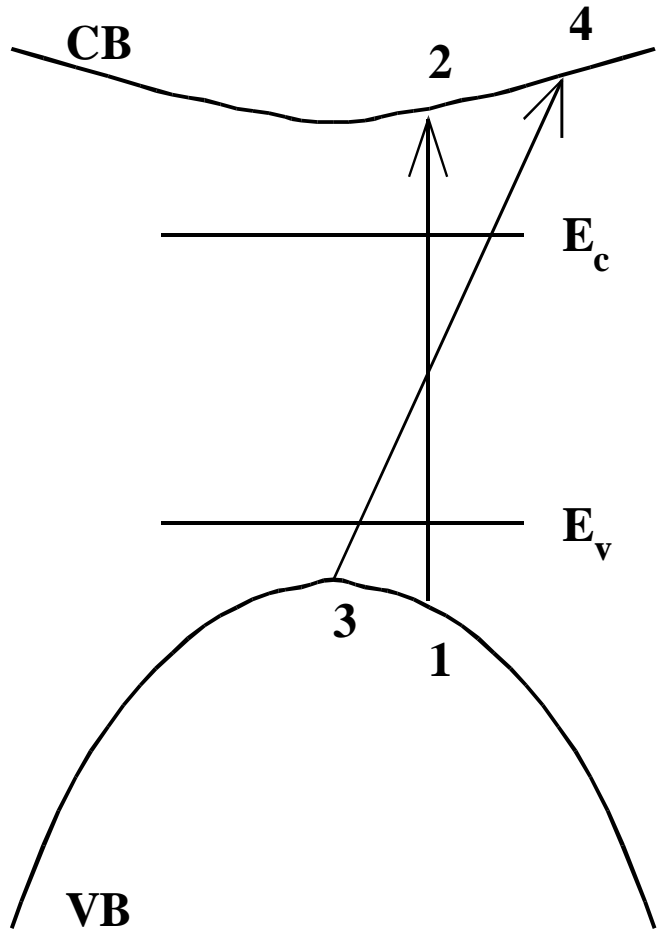,height=6.0in}}
\caption{
Schematic band structure of the direct-forbidden gap 
semiconductor.  Here $E_v$ and $E_c$ are the interface levels.  
Arrows indicate two ways that light absorption takes place:  
direct electron transition $1\rightarrow 2$, and indirect 
$3\rightarrow 4$ that occurs at the interface.  } 
\end{figure}

\begin{figure}
\centerline{\psfig{figure=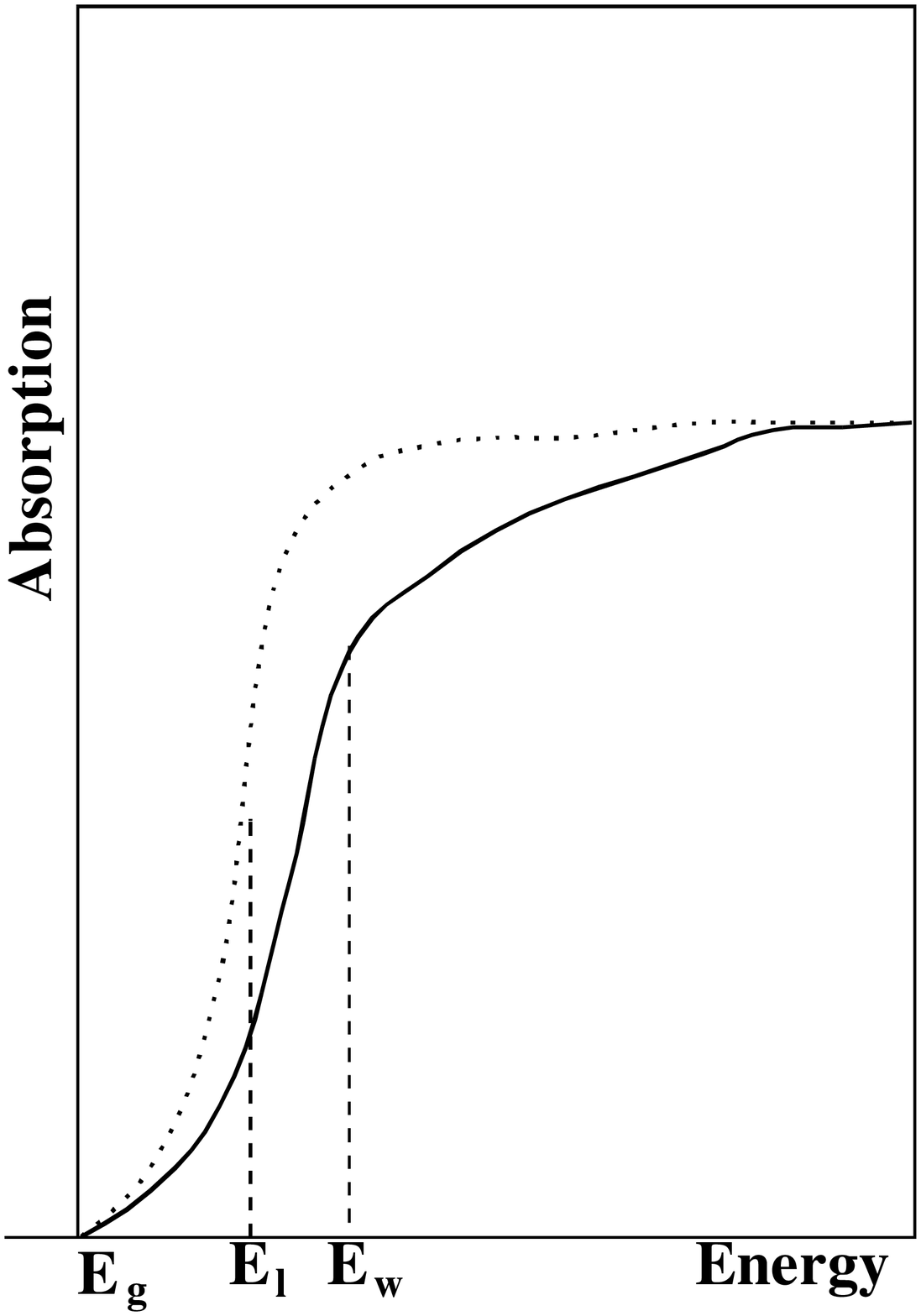,height=6.0in}}
\caption{
Interband light absorption
at the fundamental absorption edge in the
crystallite.  The dotted line indicates the influence of the
interface roughness. Here $E_l=E_g+ 
\hbar^2(2m_{c\parallel}l^2)^{-1}$, $E_w=E_g+W_c$.
}
\end{figure}

%
%

%
%


\begin{thebibliography}{99}
\bibitem{Elliot} Elliot R. J., Phys. Rev., {\bf 108},  1384 
(1957).

\bibitem{PR98}L.\ Braginsky, Phys. Rev. B, {\bf 57},  R6870 
(1998); cond-mat/9705004.

\bibitem{Ando}T.~Ando, S.~Wakahara, and H.~Akera, 
Phys.  Rev.  B, {\bf 40}, 11609   (1989). 




\bibitem{Volkov} V.\ A.\ Volkov and T.\ N.\ Pinsker, Sov. Phys. 
JETP, {\bf 72}, XXX (1977) [Z. Exp. i Teor. Fiz. {\bf 72}, 1087 
(1977)].

\bibitem{Bass}F.\ G.\ Bass and I.\ M.\ Fuks, {\it Wave 
Scattering from Statistically Rough Surfaces}, (Pergamon Press, 
1979).

\bibitem{Ivchenko}E.\ L.\ Ivchenko and G.\ E.\ Pikus, {\it 
Superlattices and Other Heterostructures. Symmetry and Optical 
Phenomena}, second ed. (Springer-Verlag, Berlin, 1997).

\bibitem{Lockwood} D.\ J.\ Lockwood, Solid State Commun., {\bf 
92}, 101 (1994); D.\ Schwall, F.\ Otter, and J.\ Galligan, 
Philosophical Magazine B, {\bf 75},  887 (1997).

\bibitem{Beck}N.\ Beck et al, J.\ of Non-Crystalline Solids,  
{\bf 198-200}, 903 (1996).



\bibitem{WO3}K.\ Glassford and J.\ Chelikowsky, Phys. Rev. B, 
{\bf 46},  1284 (1992); Shang-i Mo and W.~Y.~Ching, 
Phys.~Rev.~B, {\bf 51}, 13023 (1995).

\bibitem{Yamada} K.\ Yamada and S.\ Kohiki, Physica E, {\bf 4}, 
228 (1999).




\end{thebibliography}
\end{document}